\documentclass[twocolumn,floats,showpacs,superscriptaddress,pre]{revtex4}
\usepackage{amsmath,amssymb,bm}
\usepackage{graphics}
\usepackage{epsfig}
\usepackage{graphicx}

\begin{document}

\title{IS THE EPR PARADOX REALLY A PARADOX?}
\author{Natalia Gorobey}
\affiliation{Peter the Great Saint Petersburg Polytechnic University, Polytekhnicheskaya
29, 195251, St. Petersburg, Russia}
\author{Alexander Lukyanenko}
\email{alex.lukyan@mail.ru}
\affiliation{Peter the Great Saint Petersburg Polytechnic University, Polytekhnicheskaya
29, 195251, St. Petersburg, Russia}

\begin{abstract}
An analysis of the EPR paradox is proposed within the framework of the laws
of the dynamics of excitations of quantized fields. These excitations are
created and recorded in real time by sources and detectors (detector -
source "vice versa"). The source is defined as a region of space-time in
which the excitation amplitude of quantized fields with a certain
(acceptable!) set of quantum numbers is non-zero.
\end{abstract}

\begin{abstract}
If the excitation of a system of quantized fields with a given set of
quantum numbers does not exist, the amplitude of the "triggering" of the
source is zero. A system with a source of two photons in a singlet state and
two photon detectors is considered in detail. It is shown that the condition
of \textquotedblleft operation\textquotedblright\ of one detector with a
given direction of photon polarization uniquely determines the
\textquotedblleft adequate\textquotedblright\ choice of direction and
polarization for the second detector. This conclusion is in accordance with
the laws of motion and conservation of quantum numbers of excitations in a
system of quantized fields.
\end{abstract}

\maketitle







\section{\textbf{INTRODUCTION}}

The thought experiment proposed by Einstein, Podolsky and Rosen \cite{EPR}
was supposed to show the incompleteness of quantum mechanics, or more
precisely, the incompleteness of the description of micro-phenomena using
the wave function. The possibility of their more detailed description using
hidden classical parameters is now excluded thanks to the theoretical work
of Bell \cite{Bell} and their experimental verification in the experiments
of Aspect \cite{Asp}. However, the incompleteness of quantum mechanics is
revealed from another side -- together with the realization of the need for
secondary quantization \cite{BHJ} and the construction of quantum field
theory \cite{Dir}. Now microparticles should be understood as excitations of
quantized fields and, strictly speaking, in the description of
microphenomena one should completely switch to the language of QFT. In
relation to the EPR paradox, such a necessity is noted in the work of
Prokhorov \cite{Prokh}. Since the QFT formalism is intended to describe the
dynamics of systems with a variable number of particles in terms of the
amplitudes of the corresponding processes (elements of the S-matrix), it is
necessary to formulate the statement of the problem with a set of particles
\textquotedblleft controlled\textquotedblright\ in a given experiment. Here
we abandon the description of the state of the controlled set using a joint
wave function, since it inevitably leads to the EPR paradox, which should be
understood as an indication of the inadequacy of such a description.
Instead, this paper introduces the concept of a source (sink) of particles
that changes the excitation state of a quantized field "controllably" in
some region of space-time. This concept should not be confused with formal
sources of quantized fields in the functional formalism of QFT \cite{Col}.
Also, our approach should not be confused with Schwinger's source theory
\cite{Schw}, which attempts to abandon the quantized field as a fundamental
concept. All we need is the birth in a controlled region of space-time of a
certain admissible set of quantum numbers of particles and also its
controlled detection. The formalism of the sources is also in agreement with
the statement: \textquotedblleft a photon is a registered
photon\textquotedblright\ (see \cite{Bel}).

In the next section, a source theory is proposed. In the second section, the
concept of the wave function of excitation of quantized fields is introduced
within the framework of source theory. In the third section, the source
formalism within QFT is used to analyze the EPR experiment.

\section{SOURCE IN QUANTUM FIELD THEORY}

Let us proceed immediately to the construction of the source formalism. For
definiteness, as the initial field theory we shall consider "scalar"
electrodynamics with one complex scalar field of matter, described by the
Lagrangian density

\begin{equation}
\mathit{L}=-\frac{1}{4}F_{\mu \nu }F^{\mu \nu }+\eta ^{\mu \nu }\nabla _{\mu
}\varphi ^{\bullet }\nabla _{\nu }\varphi -V\left( \varphi \right) ,
\label{1}
\end{equation}%
where indicated:

\begin{equation}
F_{\mu \nu }=\partial _{\mu }A_{\nu }-\partial _{\nu }A_{\mu },\nabla _{\mu
}=\partial _{\mu }+ieA_{\mu },  \label{2}
\end{equation}%
and $\eta _{\mu \nu }=diag\left( +1,-1,-1-1\right) .$We need the electric
current vector:

\begin{equation}
j_{\mu }=ie\left( \nabla _{\mu }\varphi ^{\bullet }\varphi -\varphi
^{\bullet }\nabla _{\mu }\varphi \right) .  \label{3}
\end{equation}%
We will also need the energy-momentum tensor of this theory,

\begin{eqnarray}
T^{\mu \nu } &=&-F^{\mu \alpha }\left. F_{\alpha }\right. ^{\nu }  \notag \\
&&+\eta ^{\mu \alpha }\eta ^{\nu \beta }\left( \nabla _{\alpha }\varphi
^{\bullet }\nabla _{\beta }\varphi +\nabla _{\beta }\varphi ^{\bullet
}\nabla _{\alpha }\varphi \right) -\eta ^{\mu \nu }\mathit{L,}  \label{4}
\end{eqnarray}%
tensor of orbital angular momentum,

\begin{equation}
M^{\mu ,\alpha \beta }=x^{\alpha }T^{\mu \beta }-x^{\beta }T^{\mu \alpha },
\label{5}
\end{equation}%
spin momentum,

\begin{equation}
S^{\mu ,\alpha \beta }=F^{\alpha \mu }A^{\beta }-F^{\beta \mu }A^{\alpha },
\label{6}
\end{equation}%
and the total angular momentum of the system of fields:

\begin{equation}
J^{\mu ,\alpha \beta }=M^{\mu ,\alpha \beta }+S^{\mu ,\alpha \beta }.
\label{7}
\end{equation}%
All these quantities obey local conservation laws (continuity equations),
which are consequences of the equations of motion of fields, and they
determine the canonical generators of the corresponding symmetries, which
are also the main observables of the theory. Thus, the electric charge in a
certain region $\Omega _{0}$ of space,

\begin{equation}
Q=\int_{\Omega _{0}}d^{3}xj^{0},  \label{8}
\end{equation}%
serves as a canonical generator of the gradient transformation of fields in
this region. We begin the definition of the source with the differential
equation of the creation of a certain amount of charge in $\Omega _{0}$,

\begin{equation}
\partial _{\mu }j^{\mu }=D,  \label{9}
\end{equation}%
where $D$ is the spatial power density of the charge source ($%
[D]=C/m^{3}\cdot \sec $). In accordance with Eq.(\ref{9}), in the region $%
\Omega _{0}$ we have:

\begin{equation}
\frac{dQ}{dt}=\int_{\Omega _{0}}d^{3}x\partial _{0}j^{0}=-\int_{\Omega
_{0}}d^{3}x\partial _{l}j^{l}+\int_{\Omega _{0}}d^{3}xD.  \label{10}
\end{equation}%
If the source worked for a time $\tau _{0}$, and the generated charge left $%
\Omega _{0}$ during this time,

\begin{equation}
\Delta Q=0=-\int_{t_{0}}^{t_{0}+\tau _{0}}dt\int_{\Omega _{0}}d^{3}x\partial
_{l}j^{l}+\int_{t_{0}}^{t_{0}+\tau _{0}}dt\int_{\Omega }d^{3}xD.  \label{11}
\end{equation}%
Thus, the condition that in the region $\Omega _{0}$ during time $\tau _{0}$
a charge $Q$ was born (and left this region) is written in the form:

\begin{equation}
-\int_{t_{0}}^{t_{0}+\tau _{0}}dt\int_{\partial \Omega _{0}}d\sigma
_{m}j^{m}+Q=0.  \label{12}
\end{equation}%
The conditions for certain quantities of energy and momentum to be generated
in the region $\Omega _{0}$ during time $\tau _{0}$ is written down in a
completely analogous way. These quantities are canonical generators of
space-time translations. The corresponding condition has the form:

\begin{equation}
-\int_{t_{0}}^{t_{0}+\tau _{0}}dt\int_{\partial \Omega _{0}}d\sigma
_{m}T^{ma}+P^{a}=0.  \label{13}
\end{equation}%
Let us also consider in detail the birth of the angular momentum in the
source. The total angular momentum contained at a given moment in the region
$\Omega _{0}$ is equal to:

\begin{equation}
J_{i}=\int_{\Omega _{0}}d^{3}xJ^{0,kl}\varepsilon _{kli},  \label{14}
\end{equation}%
and the rate of change of the square of this vector,

\begin{equation}
\frac{d}{dt}J^{2}=-2\int_{\Omega _{0}}d^{3}xJ^{0,kl}\varepsilon
_{kli}\int_{\Omega _{0}}d^{3}y\partial _{0}J^{0,mn}\varepsilon _{mni}.
\label{15}
\end{equation}%
Therefore, the condition for the birth of the square of the angular momentum
in the region $\Omega _{0}$ during time $\tau _{0}$ is written in the form:

\begin{equation}
-2\int_{t_{0}}^{t_{0}+\tau _{0}}dt\int_{\Omega
_{0}}d^{3}xJ^{0,kl}\varepsilon _{kli}\int_{\partial \Omega _{0}}d\sigma
_{p}J^{p,mn}\varepsilon _{mni}+J^{2}=0.  \label{16}
\end{equation}%
Finally, in connection with the discussion of the EPR experiment, a special
role in this work will be played by the generation of a certain value of the
projection of the angular momentum onto the selected direction $\mathbf{a}$.
The corresponding additional condition has the form:

\begin{equation}
-\int_{t_{0}}^{t_{0}+\tau _{0}}dt\int_{\partial \Omega _{0}}d\sigma _{p}%
\frac{1}{\left\vert \mathbf{a}\right\vert }a^{i}J^{p,kl}\varepsilon
_{kli}+J_{a}=0.  \label{17}
\end{equation}%
In what follows we will consider only those classical observables that can
be measured simultaneously in quantum theory. Their values (quantum numbers)
form a complete set of observables in quantum theory that uniquely
identifies the excitation arising in the source.

By collecting together the additional conditions Eqs.(\ref{12}-\ref{17}) as
an addition to the original action of the theory in the region $\left[
t_{0},t_{0}+\tau _{0}\right] \times \Omega _{0}$ with the corresponding
Lagrange multipliers, we obtain the action of the source:

\begin{eqnarray}
I_{0}+\partial I_{0} &=&\int_{t_{0}}^{t_{0}+\tau _{0}}dt\int_{\Omega
_{0}}d^{3}x\mathit{L}  \notag \\
&&-\lambda _{Q}\int_{t_{0}}^{t_{0}+\tau _{0}}dt\int_{\Omega
_{0}}d^{3}x\partial _{l}j^{l}  \notag \\
&&-\lambda _{P^{a}}\int_{t_{0}}^{t_{0}+\tau _{0}}dt\int_{\partial \Omega
_{0}}d\sigma _{m}T^{ma}  \notag \\
&&-2\lambda _{J^{2}}\int_{t_{0}}^{t_{0}+\tau _{0}}dt\int_{\Omega
_{0}}d^{3}xJ^{0,kl}\varepsilon _{kli}  \notag \\
&&\times \int_{\partial \Omega _{0}}d\sigma _{p}J^{p,mn}\varepsilon _{mni}
\notag \\
&&-\lambda _{J_{a}}\int_{t_{0}}^{t_{0}+\tau _{0}}dt\int_{\partial \Omega
_{0}}d\sigma _{p}\frac{1}{\left\vert \mathbf{a}\right\vert }%
a^{i}J^{p,kl}\varepsilon _{kli}  \notag \\
&&+\left[ \lambda _{Q}Q+\lambda _{P^{a}}P^{a}\right.  \notag \\
&&\left. +\lambda _{J^{2}}\hslash ^{2}J\left( J+1\right) +\lambda
_{J_{a}}\hslash J_{a}\right]  \label{18}
\end{eqnarray}%
All that is contained in the additional conditions is the statement that in
the specified region of space-time the change in the complete set of
observables (admissible in quantum theory) is definite and will remain so
after quantization. Note that the components of the source power, $D$, etc.,
do not enter explicitly into the action Eq.(\ref{18}). We can quantize the
system in the region $\Theta _{0}=\left[ t_{0},t_{0}+\tau _{0}\right] \times
\Omega _{0}$ independently of the rest of space-time if the field values on
the boundary $\partial \Theta _{0}=\left[ t_{0},t_{0}+\tau _{0}\right]
\times \partial \Omega _{0}$ are considered and quantized as an independent
system. This is the structure of the action Eq.(\ref{18}) of the source. The
attachment of the source to the rest of the space-time along the boundary $%
\partial \Theta _{0}$ is carried out in the form of an obvious boundary
condition: the boundary value of the wave function of the fields in the rest
coincides with the wave function of the independent system on the boundary
formed by the boundary values of the fields.

When quantizing the source, we encounter one fundamental difficulty: the
fourth term (in the third line) of the total action Eq.(\ref{18}) contains
the interaction of the boundary and interior parts of the source. With the
canonical approach to quantization, the source Hamiltonian function is not
quadratic with respect to the canonical momenta. This means, in particular,
that the Feynman functional integral, which we are going to use in our
constructions, does not have a canonical definition \cite{FadSlav}. In order
to define the functional integral, we will use an analogue of the
Hubbard-Stratonovich formula \cite{Hab},\cite{Strat} to transform the fourth
term in Eq.(\ref{18}):

\begin{eqnarray}
&&\exp \left( -2i\lambda _{J^{2}}J_{k}\partial J_{k}\right)  \notag \\
&=&\frac{1}{\pi ^{3}}\int d^{6}q\exp \left[ -i\left( A_{\alpha \beta
}q_{\alpha }q_{\beta }+2\sqrt{\lambda _{J^{2}}}q_{\alpha }b_{\alpha }\right) %
\right] ,  \label{19}
\end{eqnarray}%
where

\begin{equation}
\widehat{A}\equiv \left(
\begin{array}{cc}
\widehat{0}_{3} & \widehat{E}_{3} \\
\widehat{E}_{3} & \widehat{0}_{3}%
\end{array}%
\right) ,\widehat{E}_{3}\equiv \left(
\begin{array}{ccc}
1 & 0 & 0 \\
0 & 1 & 0 \\
0 & 0 & 1%
\end{array}%
\right)  \label{20}
\end{equation}%
and

\begin{equation}
\widehat{b}\equiv \left(
\begin{array}{c}
J_{k} \\
\partial J_{k}%
\end{array}%
\right) ,\partial J_{k}\equiv \int_{\partial \Omega _{0}}d\sigma
_{p}J^{p,mn}\varepsilon _{mnk}.  \label{21}
\end{equation}%
Taking into account such a transformation of the fourth term in Eq.(\ref{18}%
), one can determine the functional integral for the source amplitude in the
canonical manner \cite{FadSlav}, but with additional integration over the
auxiliary variables $q_{\alpha },\alpha =1,2,3,4,5,6$:

\begin{eqnarray}
\mathit{K}_{0} &=&\frac{1}{\pi ^{3}}\int d^{6}q\int \prod\limits_{\mathit{%
\Theta }_{0}}\mathit{D}^{4}A\Delta \left( A\right) \mathit{D}\varphi ^{\ast }%
\mathit{D}\varphi  \notag \\
&&\times \prod\limits_{\partial \mathit{\Theta }_{0}}\mathit{D}^{4}A\Delta
\left( A\right) \mathit{D}\varphi ^{\ast }\mathit{D}\varphi \exp \left[
\frac{i}{\hslash }\left( I+\partial I\right) ^{\prime }\right] ,  \label{22}
\end{eqnarray}%
where $\Delta (A)$ is the Faddeev-Popov determinant \cite{FadPop}, and in
the action $\left( I+\partial I\right) ^{\prime }$ transformation Eq.(\ref%
{19}) is taken into account. To attach the source to the rest of the system
of quantized fields, the value of the boundary functional integral

\begin{equation}
\partial \mathit{K}_{0}=\int \prod\limits_{\partial \mathit{\Theta }_{0}}%
\mathit{D}^{4}A\Delta \left( A\right) \mathit{D}\varphi ^{\ast }\mathit{D}%
\varphi \exp \left[ \frac{i}{\hslash }\left( \partial I\right) ^{\prime }%
\right]  \label{23}
\end{equation}%
at each moment of time of operation of the source from the interval $\left[
t_{0},t_{0}+\tau _{0}\right] $ should be considered as boundary value of the
functional integral for the remaining part.

However, the complete inclusion of the source "into work" will occur after
fixing certain (admissible!) values of quantum numbers $Q,P^{a},J^{2},J_{%
\mathbf{a}}$ and additional integration of the functional integral over the
corresponding Lagrange multipliers $\lambda _{Q},\lambda _{P^{a}},\lambda
_{J^{2}},\lambda _{J_{\mathbf{a}}}$. If in classical theory this is achieved
by the condition of the extremum of the action Eq.(\ref{18}) (with arbitrary
values of quantum numbers) with respect to the mentioned Lagrange
multipliers, in quantum theory we must integrate them. However, the values
of quantum numbers cannot be arbitrary now. We know that in QFT they take
discrete values. And besides, some sets of quantum numbers are excluded
altogether. In the latter case, integration over Lagrange multipliers must
lead to a zero amplitude of the process. Thus, the conditions for the
operation of the source allow us to establish the spectrum of admissible
excitations in a given QFT together with their full set of quantum numbers.
In physical language, this means that in the region under consideration, a
particle interaction reaction is possible (and actually occurs), resulting
in an excitation with a given set of quantum numbers. In this connection, we
will also point out the case when a set of quantum numbers from the category
of possible ones may turn out to be unacceptable. In quantum chromodynamics,
based on the group of internal symmetry $SU(3)$, the fundamental "building
blocks" of hadrons - quarks - are not observed in a free state. This fact is
formulated as the principle of color confinement \cite{BPS}. It is
appropriate to recall here that, in contrast to electrodynamics with a
single charge $Q$ as the generator of internal symmetry, the $SU(3)$ group
generates a broader set of quantum numbers classifying hadrons. The full set
of commuting operators associated with internal symmetry includes one of the
Gell-Mann matrices, such as $\widehat{\gamma }_{8}$, as well as two Casimir
matrices - a quadratic combination of the Gell-Mann matrices, analogous to
the total angular momentum, and a cubic \cite{PP}.

\section{WAVE FUNCTION OF A PARTICLE IN SOURCE THEORY}

Here we consider for simplicity a source of a charged scalar particle
localized in the space-time region $\Theta _{0}$, with an electron charge,

\begin{equation}
e:Q=e,p_{0}=\sqrt{\mathbf{p}^{2}+m^{2}},\mathbf{a}=\mathbf{p},J=0,J_{a}=0,
\label{24}
\end{equation}%
and we will call this particle an electron. Its observed charge and mass,
generally speaking, differ from those in the Lagrange function Eq.(\ref{1}).
Let at $t=0$ the system of quantized fields be in some stationary state $%
\psi _{0}$. This may be, in particular, the vacuum state, which is defined
here, since, as we will see below, all constructions are localized to a
finite region of space-time, and the excitations of fields have the
character of wave packets. If after the excitation occurs the electron is
not registered, then the system will be in the excited state $\psi _{T}$ at
time $t=T$, and it should be expected that the scalar product $\left\langle
\psi _{T}\right\vert \left. \psi _{0}\right\rangle =0$. However, for an
electron, as for any particle, the statement is true: an electron is a
registered electron. Therefore, we will place a detector (a source "in
reverse") on the "supposed path" of the electron and try to register it. The
detector destroys the excitation of the system, or more precisely, it
transfers the quantum numbers of the particle to some reaction products. It
is important for us that the latter remain in the detector and are not taken
into account further in the dynamics of the fields. Note that in
experimental practice it is desirable to have as large as possible the
spatial-temporal dimensions (coherence parameters) of the source. In
accordance with the Heisenberg uncertainty principle, the momentum-energy
characteristics of the particle state will have a smaller spread relative to
the average values $p_{0},\mathbf{p}_{0}$. We will now determine this state
using a set of detectors. In quantum mechanics, the state of a particle
(wave function) born in a certain localized region has the character of a
wave packet, which also expands as it moves. In order to determine the shape
of the packet more accurately, we will need detectors of relatively smaller
sizes, although we will need more of them. We will place the detectors on
the forward hemisphere (relative to $\mathbf{p}_{0}$) with the center at the
source. Taking into account the main feature of quantum theory, which
consists in the fact that a particle is always registered entirely by one
detector, we obtain in the limit a set of points with coordinates on a
hemisphere -- potential localizations of an electron. In this we see the
essence of corpuscular-wave dualism: in the source the state of the particle
is closest to the plane De Broglie wave, and on the screen with detectors we
observe a reduction to a localized state.

Let us formulate this experimental situation within the framework of the
formalism of source theory. If the detector is ideal, we expect that the
electron flying towards it will be recorded and the resulting state of the
field system $\psi _{T}$, with an accuracy of a constant complex factor $%
\chi $, will return to the initial state $\psi _{0}$ without additional
excitation. Even if the detector "misses" the electron flying in its
direction, the scalar product

\begin{equation}
\left\langle \psi _{T}\right\vert \left. \psi _{0}\right\rangle =\chi
\label{25}
\end{equation}%
"cuts off" the unsuccessful outcome. In this case, the remaining excited
state $\psi _{T}$ will be orthogonal to the initial state. In the formalism
of the source theory, $\psi _{T}$ is determined by our functional integral,
which, before one of the detectors is triggered, is a function of their full
set of Lagrange multipliers. The operation of one of the detectors on the
screen, say, at a point with Cartesian coordinates $\mathbf{x}$ at time $%
t=t_{1}$ in our formalism means additional integration over the Lagrange
multipliers of the detector when substituting suitable values of the energy
and momentum of the particle $p_{0}=\sqrt{\mathbf{p}^{2}+m^{2}},\mathbf{p}$.
We get the scalar product Eq.(\ref{25}) as a function of the coordinates of
the triggered detector $\mathbf{x}$ at time $t_{1}$. Thus, the wave function
$\chi \left( t_{1},\mathbf{x}\right) $ is obtained, which describes
literally all potential possibilities of registering an electron on the
screen after its emission by the source. This is a prediction of the theory
of sources connected to the dynamics of a system of quantized fields. When
repeating such an experiment with a single electron in practice many times,
the probability distribution of hits on the screen should be proportional to
$\left\vert \chi \left( t_{1},\mathbf{x}\right) \right\vert ^{2}$.

\section{EPR EXPERIMENT}

Let us begin our discussion of the two-photon EPR experiment in the
framework of scalar quantum electrodynamics supplemented by our theory of
sources. Again, as the initial state of the system of quantized fields we
take $\psi _{0}$. At the moment $t=t_{0}$, a source localized in the region $%
\Theta _{0}$ begins to operate, in which excitation with a set of quantum
numbers:

\begin{equation}
2\gamma :Q=0,p_{0}=2\left\vert \mathbf{p}_{0}\right\vert ,\mathbf{p}%
=0,J=0,J_{a}=0  \label{26}
\end{equation}%
arises. It is easy to see that these are probably two photons: the
excitation has no charge, the total momentum is zero, but the energy is
nonzero. The state is singlet, so the projection of the angular momentum in
any direction $\mathbf{a}$ is zero. Our assumption about the nature of
excitation will be confirmed in the further course of events. The next
events will be the triggering of two detectors at moments $t=t_{1}$ and $%
t=t_{2}$, localized in the space-time regions $\Theta _{1}$ and $\Theta _{2}$%
. To reliably detect photons in this case, we need detectors with a large
cross-section located on opposite light beams.

Let's look at the sequence of actions more closely. If the source and two
detectors are connected to the system dynamics, but have not yet worked,
i.e. the integration over the corresponding sets of Lagrange multipliers has
not been carried out, we will do this now in a certain order. Let us first
"switch on" the source, integrating the amplitude over the corresponding set
of Lagrange multipliers of the source and substituting the set of quantum
numbers Eq.(\ref{26}). Let us recall that these parameters are already
included in the initial action of the system with sources Eq.(\ref{18}).
Thus, the experiment has begun. After this, we \textquotedblleft turn
on\textquotedblright\ the detectors, i.e. we integrate over the
corresponding sets of Lagrange multipliers and substitute two sets of
quantum numbers into the amplitude

\begin{equation}
\gamma _{1}:Q=0,p_{10}=\left\vert \mathbf{p}_{0}\right\vert ,\mathbf{p}%
_{0},J_{1}=\hslash ,J_{1\mathbf{a}_{1}}=?,  \label{27}
\end{equation}

\begin{equation}
\gamma _{2}:Q=0,p_{20}=\left\vert \mathbf{p}_{0}\right\vert ,-\mathbf{p}%
_{0},J_{2}=\hslash ,J_{2\mathbf{a}_{1}}=?.  \label{28}
\end{equation}%
To keep the intrigue, we left the projection directions and the projections
of the angular momentum for both photons undefined. Now let's be more
careful. Let's take $\mathbf{a}_{1}=\mathbf{p}_{0}$ for the first photon and
integrate the amplitude over the Lagrange multiplier $\lambda _{J_{1\mathbf{a%
}_{1}}}$, substituting specific value of $J_{1\mathbf{p}_{0}}=+\hslash $
(right-handed helicity). So, the first photon is "registered" in
right-handed helicity. Until the second detector "works", we have the right
to choose the direction of projection and the value of the projection of the
angular momentum. To calculate the integral over $\lambda _{J_{1\mathbf{a}%
_{1}}}$, we use the stationary phase method. Note that, as a consequence of
the additional condition Eq.(\ref{17}) and the structure of the action Eq.(%
\ref{18}), the equation for the stationary phase point is:

\begin{equation}
J_{1\mathbf{p}_{0}}+X=0,  \label{29}
\end{equation}%
where the second term is a function of the Lagrange multiplier $\lambda
_{J_{1\mathbf{a}_{1}}}$and, possibly, the remaining $\lambda _{J_{2\mathbf{a}%
_{2}}}$. The first term in Eq.(\ref{29}) is fixed by us. Solving equation
Eq.(\ref{29}) for $\lambda _{J_{1\mathbf{a}_{1}}}$, we find it as a function
of $J_{1\mathbf{p}_{0}}$, and of the remaining $\lambda _{J_{2\mathbf{a}%
_{2}}}$, and substitute this function into our functional integral $\psi
_{T} $. The experiment is not over - the second detector has not yet worked,
and we need to increase our attention even more. Based on what we know, the
second detector should be "tuned" to the same direction, since according to
the law of conservation of the projection of momentum,

\begin{equation}
J_{1\mathbf{p}}+J_{1\mathbf{p}}=0.  \label{30}
\end{equation}%
In this case, the result will be reliable: $\left\vert \chi \right\vert =1$.
When the experiment is repeated multiple times, the result will be
reproduced. In fact, we formulated the EPR paradox within the framework of
the theory of sources: how did it happen that reliable knowledge of the
polarization state of the second photon was obtained in the first detector?
Replacing the real EPR experiment with the extremum problem formulated above
and comparing Eq.(\ref{29}) and Eq.(\ref{30}), we do not see anything
paradoxical in the results of its solution.

We have come to the crucial test of our formalism in terms of its
correspondence to observations. Let us complicate the formulation of the
problem by an "inadequate" choice of the projection direction in the second
detector: let $\mathbf{a}_{2}\bot \mathbf{p}_{0}$. The experiment shows that
in this case the second detector \textit{can} work, and the second photon
will be registered. In the theory of sources this means that integrating the
amplitude over $\lambda _{J_{2\mathbf{a}_{2}}}$ and substituting $J_{2%
\mathbf{a}_{2}}=\hslash $ yields a non-zero result: the resulting value of
the functional integral $\psi _{T}\symbol{126}\psi _{0}$, but $\left\vert
\chi \right\vert =\alpha <1$. Obviously, $\alpha ^{2}$ is the probability of
registering a second photon when repeating the experiment multiple times in
the configuration under consideration. But the experiment shows that the
second detector \textit{may} not work. This means that the system of
quantized fields will remain in an excited state and $\left\langle \psi
_{T}\right\vert \left. \psi _{0}\right\rangle =0$. In this case, we simply
state that the system of fields cannot return to its original state and the
experiment cannot be completed. The theory of sources is applicable to the
description of completed experiments. We can also complete the experiment
within the framework of the theory of sources in case of failure with the
second detector, if we install a third detector after the second one, tuned
to the second orthogonal direction $\mathbf{b}_{2}\bot \mathbf{a}_{2},%
\mathbf{p}_{0}$. The third detector also has a solution within the framework
of the theory of sources, $\left\vert \chi \right\vert =\beta <1$.

Now it is clear where things are heading. After the first detector has been
"triggered", the polarization state of the second photon is determined
unambiguously: it is right-handed, like the first. In an "inadequate"
formulation of the experiment, we will discover this when it is repeated
many times, when either the second or third detector will be triggered with
the corresponding probability. In contrast to the experiment, the theory of
sources gives a complete set of parameters $\chi _{2},\chi _{2^{\prime }}$,
which determine the polarization state of the second photon.

Let's return to the previous stage of the experiment - registration of the
first photon. It should be remembered that we do not choose the value of the
photon projection on the direction of its movement (right-handed or
left-handed). We only state that this time it turned out like this -
right-handed. This result is introduced into the theory as our choice -- the
corresponding value $J_{1\mathbf{p}_{0}}=+\hslash $ is substituted into the
amplitude for the first detector. It can be said that the theory of sources
is always tied to a specific outcome of the experiment, and not to all its
potential possibilities. All potential possibilities can be considered
separately in the theory of sources.

Obviously, the first and second photons are equal and are registered
independently. Therefore, we can forget about the paradox and formulate a
theory of sources for all possible outcomes. Let the first photon be
registered by a pair of detectors tuned to two directions of polarization $%
\mathbf{a}_{1},\mathbf{b}_{1}$ orthogonal to each other (and to the momentum
$\mathbf{p}_{0}$) - two crossed polarizers. In the same way, the second
photon is detected by crossed polarizers $\mathbf{a}_{2},\mathbf{b}_{2}$.
There are four possible outcomes of the experiment. And the theory of
sources gives us four corresponding amplitudes $\chi \left( \mathbf{a}_{1},%
\mathbf{a}_{2}\right) ,\chi \left( \mathbf{a}_{1},\mathbf{b}_{2}\right)
,\chi \left( \mathbf{b}_{1},\mathbf{a}_{2}\right) ,\chi \left( \mathbf{b}%
_{1},\mathbf{b}_{2}\right) $. The squares of the moduli of these amplitudes
give the corresponding probabilities, the sum of which is equal to one. Here
we find ourselves in a situation similar to the Aspect experiment \cite{Asp}.

\section{CONCLUSIONS}

The formalism of sources (sinks) proposed in this work allows us to describe
real processes of excitation and detection of excitations of quantized
fields in a limited region of space-time, and not the amplitudes of possible
processes (elements of the scattering matrix). These sources (sinks)
themselves occupy limited regions in space-time, so that the dynamics of the
excited system of quantized fields obeys the standard laws (the Schr\"{o}%
dinger equation and conservation laws) outside these devices. There is no
doubt that an entangled pair of photons produced by a source in a singlet
state will "behave" in accordance with relativistic causality and
conservation laws when subsequently registered. The formalism of sources
(sinks) gives us the right to choose the direction of the projection of the
moment when registering one of the photons. The "adequate" behavior of the
second photon is predetermined by the laws of motion of the system of
quantized fields, and not by superluminal speeds of information transfer.
Taking into account the laws of motion of the QFT, we do not see anything
unusual in the results of the detectors' work.

To summarize the entire discussion, we emphasize that our conclusions are
based on the balance of quantum numbers for excitations of quantized fields:
the equality to zero of the projection of the angular momentum in any
direction in the source is \textquotedblleft transmitted\textquotedblright\
(conservation laws) to the detectors in accordance with the principle of
causality. But the structure of the detector is such (see additional
conditions Eq.(\ref{17})) that its \textquotedblleft switching
on\textquotedblright\ in the mode of projection of the photon moment onto
any direction chosen by us, uniquely determines this projection for the
second photon as well. After this, the choice of the design direction and
the measurement result in the second detector are predetermined. Everything
is determined by the observer who will be the first to integrate the
amplitude over the set of Lagrange multipliers available to him!

\section{ACKNOWLEDGEMENTS}

We are thanks V.A. Franke for useful discussions.




\end{document}